\documentclass[12pt,fleqn]{article}
\usepackage{ifthen}                   % needed for Feyman-Slash macro
\usepackage{exscale}                  % correct scaling of math-symbols
\usepackage[intlimits]{amsmath}       % improved mathematical typesetting
\usepackage{amsfonts}
\usepackage{amssymb,amscd}
\usepackage[dvips]{epsfig}                   
\usepackage{array}
\usepackage{wrapfig}
\usepackage{graphics}

\begin{document}

\psfull

{\small\rightline{arXiv:hep-ph/0411060}}

\vspace{1cm}

%Partial derivatives
\newcommand*{\pd}{\partial}
\newcommand*{\pdm}{\pd_{\mu}}
\newcommand*{\pdn}{\pd_{\nu}}
\newcommand*{\pdi}{\pd_{i}}
\newcommand*{\pda}[1]{\pd_#1}
\newcommand*{\mn}{{\mu\nu}}

%Convenient formatting

\newcommand*{\be}{\begin{equation}}
\newcommand*{\ee}{\end{equation}}

\newcommand*{\bea}{\begin{eqnarray}}
\newcommand*{\eea}{\end{eqnarray}}

\newcommand*{\pref}[1]{(\ref{#1})}

\newcommand*{\prefr}[2]{(\ref{#1}-\ref{#2})}

\newcommand*{\img}{\mathrm{im}}
\newcommand*{\rr}{\mathbb{R}}
\newcommand*{\brst}{\mathrm{BRST}}
\newcommand*{\tl}{\mathrm{tl}}

\begin{center}
\begin{LARGE}
{\bf The Ghost-Gluon Vertex\\ in Landau Gauge Yang-Mills Theory$^*$}\\
\end{LARGE}
\vspace{1.2cm}
{\bf  W.\ Schleifenbaum$^{a}$, A.\ Maas$^{a}$, J.\ Wambach$^{a,b}$,
and R. Alkofer$^{c}$\\}
\vspace{0.3cm}
{$^a$Institute for Nuclear Physics, Technical University Darmstadt,\\
Schlo{\ss}gartenstr.\ 9, D-64289 Darmstadt, Germany\\ }
\vspace{0.3cm}
{$^b$Gesellschaft f\"ur Schwerionenforschung mbH,\\ Planckstr. 1, 
D-64291 Darmstadt, Germany\\}
\vspace{0.3cm}
{$^c$Institute for Theoretical Physics, T\"ubingen University,\\
Auf der Morgenstelle 14, D-72076 T\"ubingen, Germany\\}
\vspace{0.5cm}
{$^*$ Talk given by W.S. at the International School of Subnuclear Physics,\\
August 29 - September 07, 2004, in Erice, Italy.}
\end{center}
\vspace{0.5cm}

\begin{abstract}
\noindent 
Possible mechanisms for gluon confinement in Yang-Mills theory in covariant
gauges are discussed briefly. Implications for the infrared behavior of the
Green's functions are outlined. Results for the gluon and Faddeev-Popov ghost
propagators from Dyson-Schwinger studies and lattice calculations are
presented. The importance of the ghost-gluon vertex in these studies is
discussed. Numerical results for a non-perturbatively constructed ghost-gluon
vertex in four and three space-time dimensions are presented.
\end{abstract}
\vspace{0.5cm}

\section{Confinement in Covariant Gauges}

The strong interactions governing hadron and ultimately nuclear dynamics are
described by quantum chromodynamics, the theory of quarks and gluons. This
theory has several genuine non-perturbative features, such as {\it e.g.\/}
dynamical chiral symmetry breaking. The most prominent phenomenon is, however,
confinement. Strong evidence exists that it is generated in the Yang-Mills
subsector, {\it i.e.\/} by the gluons.
Na\"ively, a Yang-Mills theory is defined in terms of the generating functional
\cite{Rivers:1987hi}
\bea
Z[J]&=&\int{\cal D}A_\mu^a\; \exp\left(-\int d^dx \; {\cal L} \; + \; \int d^dx \;
J_{\mu}^a A_\mu^a \right) \; , \label{part}\\
{\cal L}&=&\frac{1}{4}F_{\mu\nu}^aF_{\mu\nu}^a\; ,\nonumber\\
F^a_{\mu\nu}&=&\pdm A_\nu^a-\pdn A_\mu^a+g_df^{abc}A_\mu^bA_\nu^c\nonumber.
\eea
Here we have already chosen a description in  Euclidean space-time, which is
used throughout this work. $A_\mu^a(x)$ is the gluon field, $F^a_\mn(x)$ is the field
strength tensor, $g_d$ is the gauge coupling constant in $d$ dimensions, and
$f^{abc}$ are the structure constants of the gauge group. The measure of the
functional integral \pref{part} extends over the complete configuration 
space of the gluon field.

However, the functional integral \pref{part} is ill-defined. The 
Lagrangian is invariant under local gauge transformations and therefore 
the complete configuration space contains infinite sets of mutually equivalent 
field configurations. They are said to lie on the same gauge orbit. To avoid 
this over-counting in the functional integral, it is necessary to 
select only one copy on each gauge orbit.  One tries to achieve this by 
gauge-fixing \cite{Rivers:1987hi}. Implementing the linear covariant gauge
fixing condition $\partial^\mu A_\mu^a=0$ leads to the generating functional
\bea
Z[J]&=&\int{\cal D}A_\mu^a \; M \; \exp\left(-\int d^dx\;  \left({\cal L}+
{\cal L}_{\mathrm{gf}} \right) \; + \; \int d^dx
J_{\mu}^a A_\mu^a \right) \; , \nonumber\\
M&=&\det(-\pdm D_\mu^{ab})\; , \label{fpdet}\\
D^{ab}_\mu&=&\pdm\delta^{ab}+g_df^{abc}A_\mu^c\; ,\nonumber
\eea
where $M$ is the Faddeev-Popov determinant, and ${\cal L}_{\mathrm{gf}}$ 
the gauge-fixing term. The determinant is then rewritten as a functional 
integral over scalar Grassmann fields, the Faddeev-Popov ghosts. 
The generating functional then reads
\bea
Z[J=0]=\int{\cal D}A_\mu^a{\cal D}\bar c^b{\cal D}c^c\; 
\exp\left(-\int d^dx  \left({\cal L}+{\cal L}_{\mathrm{gf}}
+\bar c^d\pdn D_\nu^{de}c^e\right)
 \right) \; .
\eea
The ghost and anti-ghost fields, $c^a(x)$ and $\bar c^a(x)$, respectively, 
encode part of the quantum fluctuations of the gluon field. Their appearance
is a generic feature  of covariant gauges. As will be detailed below, this
procedure is in general not sufficient to completely specify the gauge beyond the
use in perturbation theory.

One of the most intriguing features of QCD is confinement. 
Mechanisms for gluon confinement have been related to the infrared behaviour 
of propagators, see {\it e.g.\/} ref.\ \cite{Alkofer:2000wg}.
We will focus here on aspects which have been substantiated recently, {\it
cf.\/} section \ref{sprop}.

First we discuss a sufficient but not necessary condition for confinement. 
If a propagator $D$ vanishes sufficiently fast in the infrared,
\be
\lim_{p^2\to 0} \; D(p^2)=0\; ,\label{oehme}
\ee
then the ``particle'' represented by $D$ cannot have a positive  semi-definite
spectral norm. Therefore it violates the Osterwalder-Schrader axiom of
reflection positivity \cite{Osterwalder:dx}, and  the propagator $D$ cannot
have a K\"allen-Lehmann representation. Thus this ``particle'' is not part of 
the physical spectrum, and is hence confined \cite{Oehme:bj}.

For a massless excitation such as the gluon this confinement criterion is rather
intuitive. Then, the condition \pref{oehme} is just the statement that the
on-shell propagator vanishes. Therefore, such a ``particle'' does not propagate,
{\it i.e.\/} it is confined.

The criterion \pref{oehme} is related to a scenario which was introduced by 
Kugo and Ojima \cite{Kugo:gm}. In perturbation theory, ghosts and longitudinal
gluons are not part of the physical spectrum, since they form doublets with
respect to BRST charge. As these doublets are conjugate pairs by ghost number,
this is called the quartet mechanism. The conjecture of Kugo and Ojima states
that also transverse gluons form together with ghost-gluon bound states a
quartet, and are thus not part of the physical spectrum. This does not imply
that transverse gluons do not have asymptotic states. However, these states are
not in the physical subspace and therefore do not contribute to $S$-matrix
elements.

This scenario is based on a  proof which requires three preconditions. The
first is an unbroken BRST charge. Whether this is the case is not known. The
second is the failure of cluster decomposition in the complete state space.
This requires the existence of a massless excitation in the unphysical
subsector. However, at the same time a mass gap in the physical subsector is
required. The third precondition is an unbroken global color charge. This
condition can be cast in Landau gauge into a condition for the ghost propagator
$D_G$ \cite{Kugo:1995km}
\bea
\lim_{p^2\to 0} p^2 D_G(p^2)\to \infty \; .\label{zhorizon}
\eea
As will be seen below this scenario also entails condition
\pref{oehme} for the gluon propagator.

Another scenario has been proposed by Zwanziger \cite{Zwanziger:2003cf} based
on work by Gribov \cite{Gribov:1977wm}. Only one gauge copy on each gauge orbit
should be included in the functional integral constituting the generating
functional. However, no local gauge condition is known which is able to
specify the gauge completely also for large gauge field fluctuations. This is
known as the Gribov problem \cite{Gribov:1977wm}. The origin of this problem
actually provides a possible handle to understand confinement. Therefore a
closer look at the properties of the field configuration space is necessary.

It is possible to partition configuration space by the zeros of the
Faddeev-Popov determinant \pref{fpdet}. Concentric regions are found and each
gauge orbit intersects each region at least once \cite{Gribov:1977wm}. The
region containing the origin, and thus relevant for perturbation theory, is
known as the first Gribov horizon. It can be shown to be compact and convex. It
has been found, that this region also contains gauge copies, see {\it e.g.\/}
\cite{vanBaal:1997gu}, and it is necessary to further restrict to the so-called
fundamental modular region. This region is specified by minimizing an
appropriate functional along each gauge orbit \cite{Zwanziger:1993dh}. The
fundamental modular region is also convex and compact around the origin
\cite{Zwanziger:2003cf}.

The latter properties lead to an interesting observation. As configuration space
is infinite-dimensional, any volume is dominated by its boundary. It has
therefore been argued that only the common boundary of the first Gribov horizon
and the fundamental modular region contribute to any correlation function built
from a finite number of field operators, especially Green's functions
\cite{Zwanziger:2003cf}. Therefore, restricting to the first Gribov horizon,
would be sufficient to calculate Green's functions.

This has a further important consequence \cite{Zwanziger:2003cf}. The
Faddeev-Popov determinant \pref{fpdet} vanishes by definition on this boundary.
Therefore, the condition \pref{zhorizon}, termed in this context the Zwanziger
horizon condition, is fulfilled as the determinant is directly linked to the
inverse self-energy of the ghost. This also
entails that the action is dominated by the gauge fixing term alone. Thus, in
the extreme infrared all Green's functions are determined by contributions
involving ghosts. This leads to the notion of ghost dominance.

Assuming the horizon condition \pref{zhorizon}, it can be shown that an infrared
tree-level ghost-gluon vertex complies with the renormalization group in
leading order. This leads to the  hypothesis of a bare ghost-gluon vertex also in
the infrared. It has then been shown that, for the Landau gauge gluon propagator,
the condition \pref{oehme} follows and gluons are confined
\cite{Zwanziger:2003cf}.

Whether the identical infrared behavior of the ghost and gluon propagators
within the Kugo-Ojima and the Zwanziger-Gribov scenarios, respectively, is
more than a formal coincidence is not known. However, a ghost propagator with
a stronger infrared divergence than a massless pole is also intuitively seen to
be able to mediate confinement. After Fourier transformation such a singularity
corresponds to a correlation being more long-range than that of a photon, and
thus mediates a stronger force than a Coulomb force.

\section{Propagators in Yang-Mills Theory}\label{sprop}

\begin{figure} 
\epsfig{file=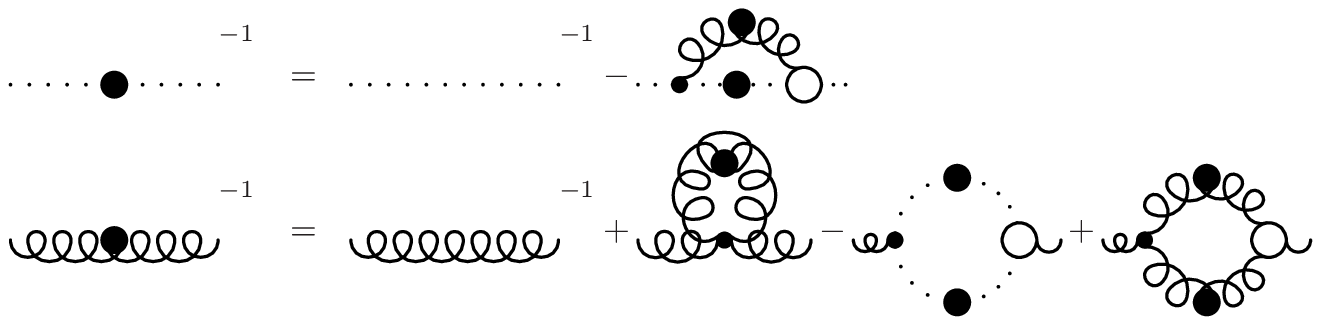,width=0.9\linewidth}
\caption{The
Dyson-Schwinger equations for the 2-point functions of Yang-Mills theory.
Dotted lines are ghosts and wiggly lines are gluons.  Lines with a dot are full
propagators, vertices with a small black dot are bare vertices and white
circles are full vertices. The equation for the gluon has already been
truncated to the one-loop level, see text.}
\label{figt0sys}
\end{figure}

In both confinement scenarios described above the infrared behavior of 
propagators plays a central role. As infrared singularities are anticipated,
{\it cf.\/} eq.\ \pref{zhorizon}, a non-perturbative method formulated in
space-time continuum is desirable. 
Such a formulation is provided by the Dyson-Schwinger equations
(DSEs), determining the Green's function of a theory, see {\it e.g.\/}
\cite{Rivers:1987hi,Alkofer:2000wg}. A graphical representation of the
equations for the propagators is shown in fig.~\ref{figt0sys}.

The Landau gauge will be used throughout, as it is advantageous in the study of
confinement for various reasons. First, the Landau gauge is a fixed point of the
renormalization group. Second, the conditions for the 
Kugo-Ojima and Zwanziger-Gribov scenario take the simple forms \pref{oehme} 
and \pref{zhorizon} in Landau gauge. Last but not least, 
it is less singular than other gauges.
This manifests itself in the bareness of the ghost-gluon vertex for a vanishing
incoming ghost momentum \cite{Taylor:ff}
\be
\lim _{p\to 0} \Gamma^{abc}_\mu(k;q,p) =  i g_d f^{abc} q_\mu.\label{taylor}
\ee
Here $p$, $q$, and $k$ are the momenta of the incoming ghost, the outgoing
ghost, and the gluon, respectively. 

Employing the ghost DSE and quite general assumptions, it can then be shown
\cite{Watson:2001yv} that the ghost propagator necessarily fulfills condition
\pref{zhorizon}. It is not possible in this generality to show that the gluon
propagator satisfies \pref{oehme}. This requires a solution of the DSEs.
However, it is not possible to solve these equations as they form an
infinite set of coupled non-linear integral equations. Therefore, it is
necessary to truncate the system.

\begin{figure}[ht]

\epsfig{file=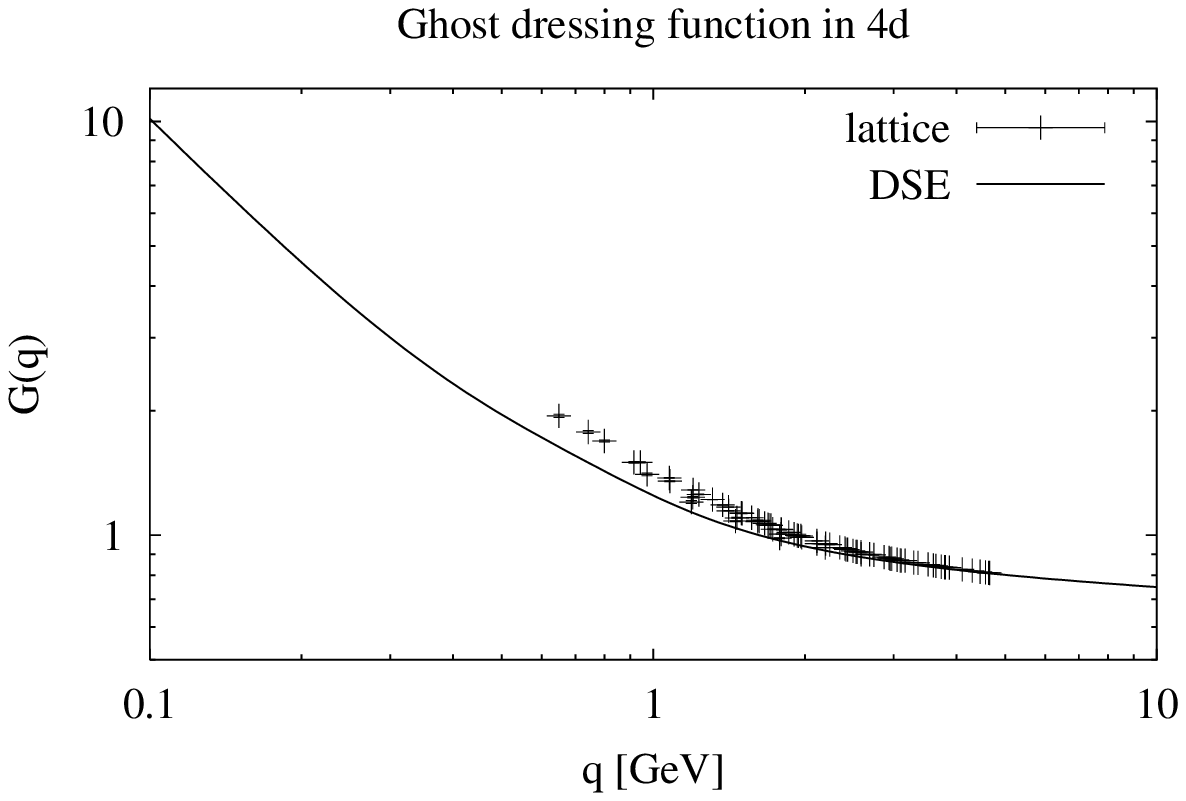,width=0.5\linewidth}\epsfig{file=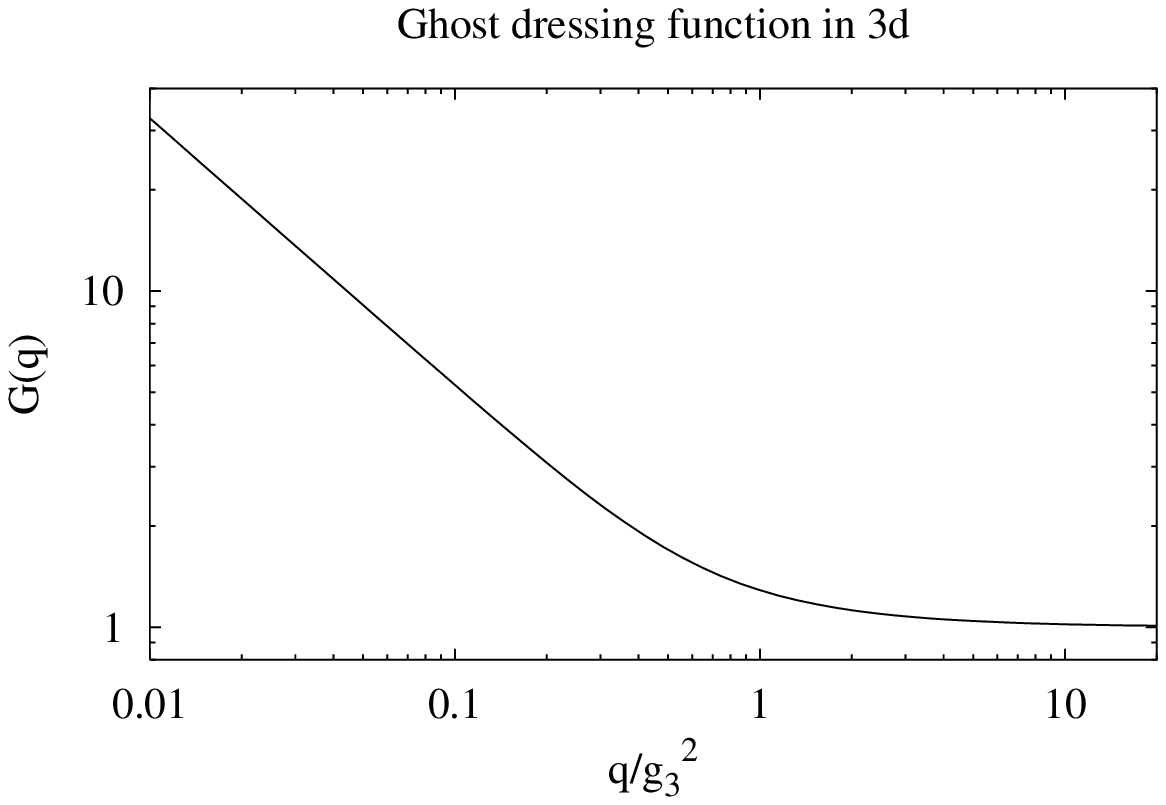,width=0.5\linewidth}
\caption{The ghost dressing function from DSEs in four dimensions 
\cite{Fischer:2002hn}
compared to lattice results \cite{Langfeld:2002dd} (left) and from DSEs in 
three dimensions
\cite{Maas:2004se} (right). $g_3$ is the dimensionful coupling constant in 
three-dimensional Yang-Mills theory.}
\label{ghostfig}
\end{figure}

\begin{figure}[ht]
\epsfig{file=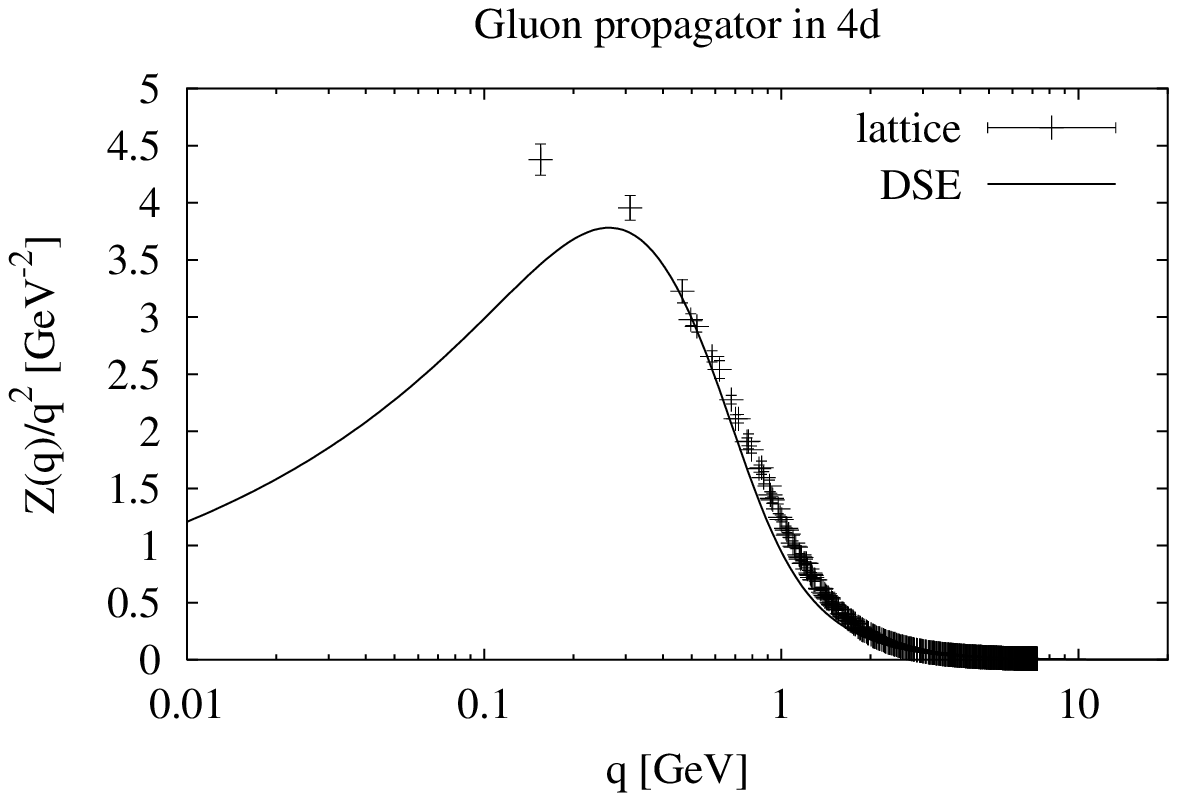,width=0.5\linewidth}\epsfig{file=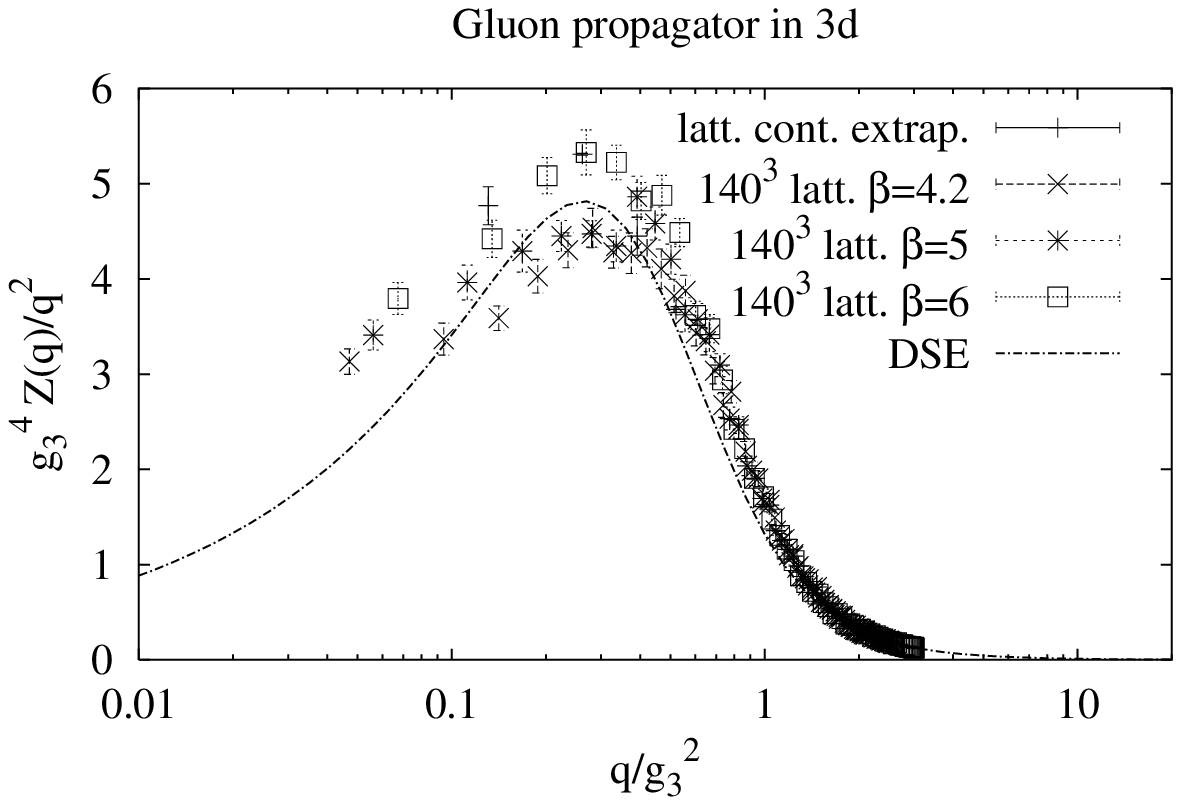,width=0.5\linewidth}
\caption{The gluon propagator from DSEs in four dimensions 
\cite{Fischer:2002hn} compared to
lattice data \cite{Bowman:2004jm} (left) and from DSEs in three dimensions 
\cite{Maas:2004se}
compared to corresponding lattice data \cite{Cucchieri:2001tw,Cucchieri:2003di}
(right). Note that the lattice data at low momenta are likely 
to be strongly influenced by finite volume effects.}
\label{gluonfig}
\end{figure}

During the last years, the truncation scheme depicted in fig.~\ref{figt0sys} has
been established \cite{vonSmekal:1997is}. This scheme has been successfully applied
to Yang-Mills theory in four dimensions \cite{Fischer:2002hn}, three dimensions
\cite{Maas:2004se} and at finite temperature \cite{Gruter:2004bb} as well as to
QCD including dynamical quarks \cite{Fischer:2003rp}. In this scheme, all
equations beyond those of the propagators are neglected, as well as genuine
two-loop contributions in the equations for the propagators. Furthermore, a
perturbative color structure of the Green's functions is assumed. It then only
remains to specify the two remaining full vertices, the ghost-gluon  and the
3-gluon vertex. The 3-gluon vertex is chosen such that leading-order resummed
perturbation theory is reproduced in the ultraviolet. Due to the property
\pref{taylor} the ghost-gluon vertex is usually taken to be bare. The results
for the ghost and gluon propagators are shown in figs.\ \ref{ghostfig} and
\ref{gluonfig} compared to lattice calculations. Both methods agree remarkably
well. The largest deviations are found at intermediate momenta
where the truncation scheme is expected to be least reliable. Although this
scheme imposes several problems, especially concerning gauge invariance,
in a systematic analysis only small quantitative effects due to
truncation errors have been found, for a detailed discussion of this issue
see {\it e.g.} refs.\ \cite{Fischer:2002hn,Maas:2004se}.

The gluon propagator obeys condition \pref{oehme}, and the gluon is thus
confined. As the ghost propagator also satisfies the Zwanziger horizon
condition \pref{zhorizon}, the Kugo-Ojima and the Zwanziger-Gribov scenarios are
both fulfilled. These findings have also been confirmed recently by exact
renormalization group methods \cite{Pawlowski:2003hq}. Furthermore, such
propagators are in sharp contrast to the older concept of infrared slavery,
which expected a gluon dressing function diverging as  $1/p^2$
\cite{Brown:1988bm}. These earlier attempts neglected the ghost contribution
\cite{Mandelstam:1979xd}. This so-called Mandelstam approximation leads to
results contradicting current lattice results and can now be dismissed.

In the infrared, ghost dominance is found. Thus, besides the full gluon
propagator, only diagrams with at least one ghost line in fig.\ \ref{figt0sys}
contribute. This is in accordance with the Zwanziger-Gribov scenario. The
single crucial point in this derivation is then the structure of the
ghost-gluon vertex \cite{Lerche:2002ep}. It is therefore of great importance to
understand it. The infrared behavior of the propagators has been studied using
several ans\"atze, motivated by the Slavnov-Taylor identity (STIs) for the
ghost-gluon vertex. The results are found to depend only quantitatively on the
input vertex \cite{Lerche:2002ep}. However, a more extensive analysis is
highly desirable and will be presented below.

\section{Ghost-Gluon Vertex}

\begin{figure}[b]
\epsfig{file=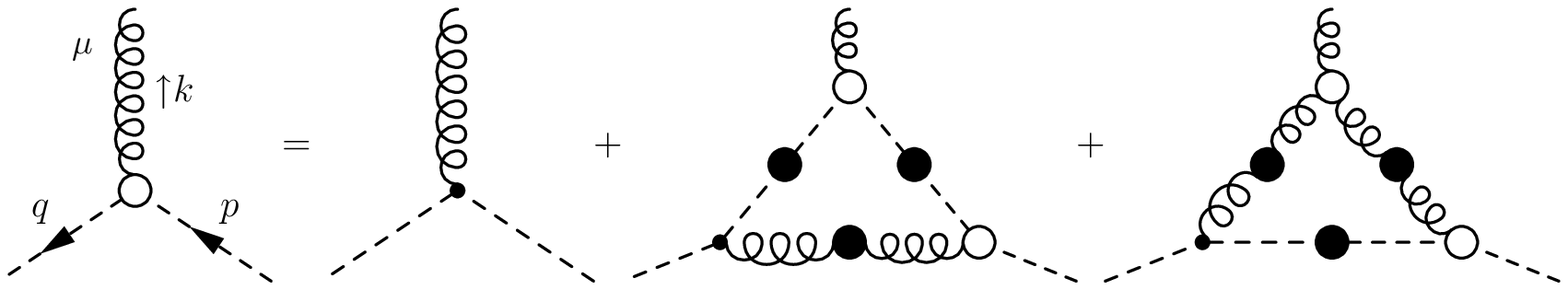,width=0.9\linewidth}
\caption{The DSE for the ghost-gluon vertex. 
Dotted lines are ghosts and wiggly lines are gluons. Lines with a dot are full
propagators, vertices with a small black dot are bare vertices and white
circles are full vertices. Contributions from the ghost-gluon scattering kernel
have been removed, see text.}
\label{figvertex}
\end{figure}

The most general tensor structure of the ghost-gluon vertex in Landau gauge 
is given by
\be
\Gamma^{abc}_\mu(k;q,p)=ig_d\left(q_\mu \left(f^{abc}+A^{abc}(k;q,p)\right)+
k_\mu B^{abc}(k;q,p)\right)\; ,
\ee
where $A^{abc}$ and $B^{abc}$ are scalar functions. 
The tree-level contribution is left explicit. As there is no indication for a
color structure different from perturbation theory~\cite{Alkofer:2000wg}, the
structure constant is factored out, $A^{abc}=:f^{abc}A$ and $B^{abc}=:f^{abc}B$.
Due to the transversality of the gluon propagator, the on-shell ghost-gluon
vertex is transverse with respect to the gluon momentum. Therefore, the function
$B$ is only relevant when not contracting with a transverse projector, {\it
e.g.\/} in the corresponding STI. Hence, only the function 
$A$ will be discussed here. A detailed account on $B$ is given elsewhere 
\cite{Schleifenbaum}. 

In principle, the ghost-gluon vertex should be determined self-consistently by
solving the DSEs for the propagators and the vertex simultaneously. A
truncated  DSE for the vertex neglecting four-point functions is shown in fig.\
\ref{figvertex}. The completely self-consistent solution of such an equation is
of significant technical complexity and has not been achieved yet. However,
due to the hypothesis of Zwanziger, it is most likely sufficient to perform a
semi-perturbative calculation, {\it i.e.\/} to do one iteration step in the
ghost-gluon vertex DSE. If the hypothesis is correct then the resulting vertex 
should not significantly deviate from the input tree-level vertex.

\begin{figure}
\centering
\epsfig{file=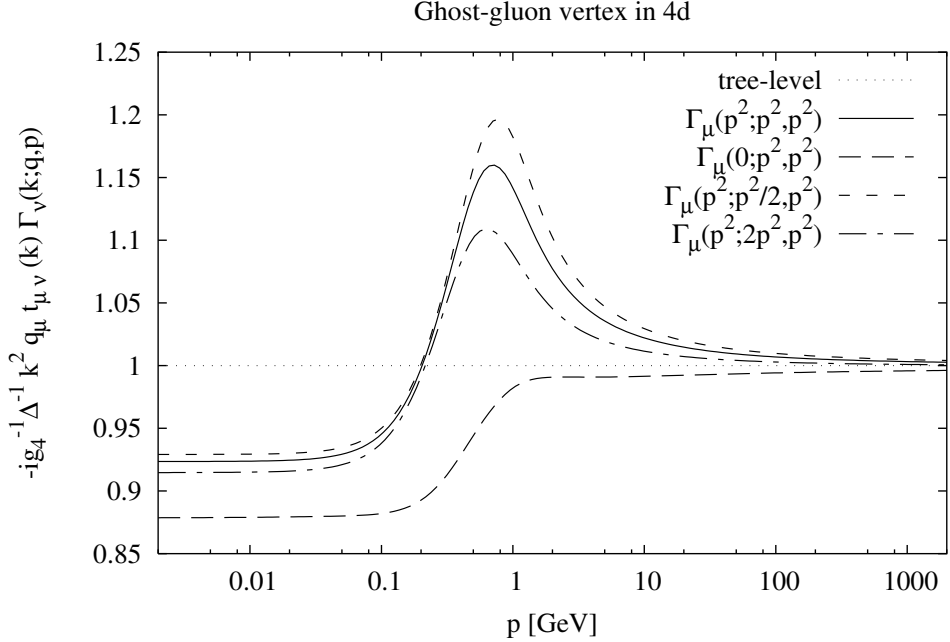,width=0.80\linewidth}
\caption{
The normalized transverse part \pref{ggvtrans} of the ghost-gluon vertex in four 
dimensions in various
kinematical regions.}
\label{fig4d}
\end{figure}

\begin{figure}
\centering
\epsfig{file=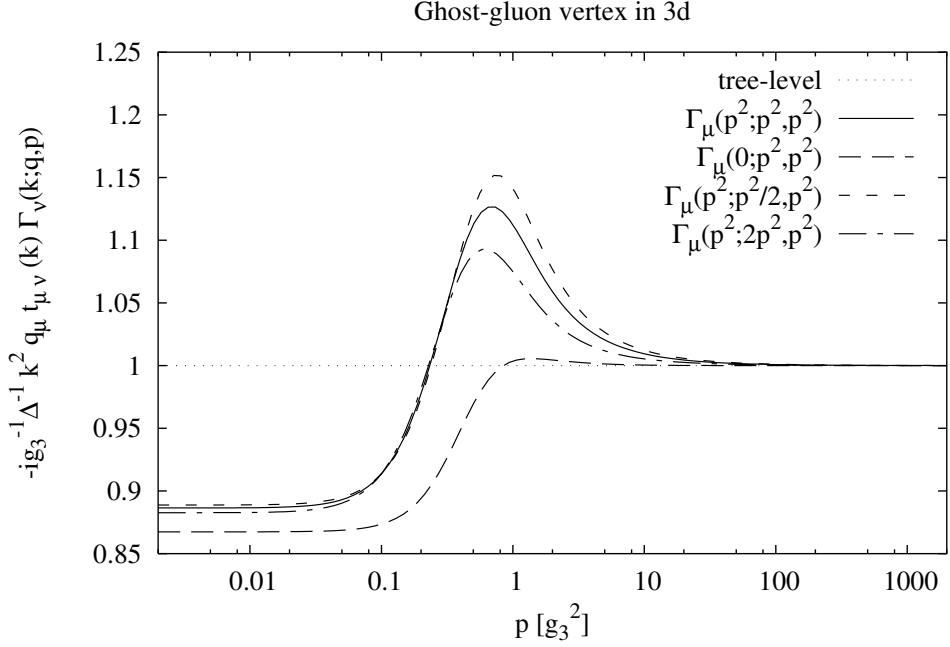,width=0.80\linewidth}
\caption{The normalized transverse part \pref{ggvtrans} of the ghost-gluon vertex 
in three dimensions
in various kinematical regions. }
\label{fig3d}
\end{figure}

Hence, the evaluation scheme neglects the ghost-gluon scattering kernel in the
ghost-gluon vertex DSE. Further, as this is the first iteration step, all
vertices are left bare and only the propagators are dressed with the results
presented in the previous section. The results are displayed in fig.\
\ref{fig4d} for four dimensions and in fig.\ \ref{fig3d} for three dimensions. 
Shown is the transverse part of the ghost-gluon vertex, $1+A$, defined by
\bea
\frac{k^2}{ig_d\Delta}q_\nu\left(\delta_\mn-\frac{k_\mu k_\nu}{k^2}\right)
\Gamma^{abc}_\mu(k;q,p) = f^{abc} (
1+A(k;q,p) )\label{ggvtrans}
\eea
where $\Delta = q^2 k^2-(q\cdot k)^2$ is a Gram determinant.

Several observations can be made \cite{Schleifenbaum}. First, although not shown in the plots, the
results comply with the infrared limit \pref{taylor} and  thus satisfy the
respective STI in this specific kinematical limit. Second, they have the proper
ghost-antighost symmetry \cite{Lerche:2002ep},  $A(k;q,p)=A(k;p,q)$. Therefore,
they fulfill all required conditions. In addition, in three dimensions,
asymptotic freedom is manifest and the vertex reduces to the bare one in
the ultraviolet. However, in four dimensions the anomalous dimensions is
incorrect, as the vertex is not resummed yet.

As can be seen, the result is a vertex which is very close to the tree-level one  
at all momenta, especially in the infrared. This does not change in other kinematical
conditions \cite{Schleifenbaum} and thus confirms Zwanziger's hypothesis
in an impressive manner. The origin for the small deviations 
from the exact bare vertex in the infrared may be twofold. On the one hand, certainly iteration
to self-consistency and truncation artifacts change the results quantitatively.
As the ghost-gluon vertex DSE is already quite well fulfilled,  this will
most likely not lead to a qualitative change. On the other hand, in 
Zwanziger's hypothesis only agreement with leading-order renormalization group
has been required. Subleading effects may generate a deviation of the order
observed.

In addition, the vertex is nearly tree-level for all momenta. This supports
the assumption of a bare ghost-gluon vertex in the truncation scheme for the
propagators. It especially provides additional evidence for ghost dominance and
therefore for the Zwanziger-Gribov confinement mechanism.

Similar results have also been reported recently in lattice calculations in 
four dimensions
\cite{Cucchieri:2004sq}. They confirm the results found independently here and
indicate also a nearly tree-level ghost-gluon vertex.

To assess the error of the approximations adopted, the full ghost-gluon
vertices in the ansatz shown in fig.\ \ref{figvertex} have been replaced by
the vertices motivated by the corresponding STI \cite{Lerche:2002ep}. It has
been found \cite{Schleifenbaum} that also these
vertices are mapped to a nearly tree-level like vertex by the truncated
DSE, as long as no singularity in an external momentum is introduced.
(As there is no iteration in the current ansatz, these
divergencies would survive the calculation.) Moreover, the results have been found
to be slightly more sensitive to the propagators than to the input vertices. 
All these variations are quantitatively small. 
This indicates that the solution of the
ghost-gluon vertex DSE is a vertex which is nearly tree-level.

It is probable that the influence of quarks, {\it cf.\/} ref.\ 
\cite{Fischer:2003rp}, or of
an adjoint Higgs field, {\it cf.\/} ref.\ \cite{Maas:2004se}, 
on these findings is
negligible. Neither couple directly to the ghost and therefore their presence
does not give rise to new terms in the ghost-gluon vertex DSE. Hence, they can
only contribute indirectly via their effect on the Yang-Mills propagators,
which are very robust against coupling a small number of quark flavours \cite{Fischer:2003rp}, or
scalars \cite{Maas:2004se}.

The ghost-gluon vertex may also be interesting with respect to extracting the
running coupling constant \cite{Cucchieri:2004sq}. This issue has to be
investigated further.

\section{Conclusions}

In Landau-gauge Yang-Mills theory we have calculated the ghost-gluon vertex
employing non-perturbative propagators. The results show quite conclusively
that deviations from the tree-level value are very small, even in the infrared.
On the one hand, this supports the truncation scheme used to calculate the
propagators of the Yang-Mills theory. The results found in these calculations give
strong evidence for a confinement mechanism of the Kugo-Ojima or
Zwanziger-Gribov type. Especially, no sign of an infrared enhanced gluon
propagator in Landau gauge is found. More importantly, the results on the
ghost-gluon vertex alone, provide independent support for the
Zwanziger-Gribov scenario as they fulfill Zwanziger's hypothesis of a bare
ghost-gluon vertex in the infrared. The results presented here are thus a
further indication for a confinement mechanism driven by the dynamics on or
near the Gribov horizon. This coincides with recent findings in Coulomb gauge
\cite{Greensite:2004ur}. All these evidences combined promise a new era of
quantitative confinement physics.

\subsection*{Acknowledgments}

W.~S.\ thanks the organizers of the International School of Subnuclear Physics 
for the very interesting meeting and for the opportunity to present this talk. 
We are grateful to C.~S.~Fischer, J.~M.~Pawlowski and L.~von~Smekal 
for helpful discussions. This work is supported by the BMBF under grant numbers 06DA917 and 06DA116, and by the Helmholtz 
association (Virtual Theory Institute VH-VI-041).

%\small

\end{document}